\documentclass{PoS}

\def\beq{\begin{equation}}
\def\eeq{\end{equation}}
\def\bea{\begin{eqnarray}}
\def\eea{\end{eqnarray}}

\def\<{\langle}
\def\>{\rangle}

\title{Strange and charmed baryons using $N_f=2$ twisted mass QCD}

\ShortTitle{Strange and charmed baryons from $N_f=2$ tmQCD}

\author{\speaker{Mauro Papinutto},Jaume Carbonell\\
        Laboratoire de Physique Subatomique et de Cosmologie,\\
UJF/CNRS-IN2P3/INPG, 53 rue des Martyrs, F-38026 Grenoble, France\\
        E-mail: \email{mauro.papinutto@lpsc.in2p3.fr,carbonel@lpsc.in2p3.fr}}

\author{Vincent Drach\\
        NIC, DESY, Platanenallee 6, D-15738 Zeuthen, Germany\\
        E-mail: \email{vincent.drach@desy.de}}

\author{Constantia Alexandrou\\
        Department of Physics, University of Cyprus, P.O. Box 20537, 1678
        Nicosia, Cyprus\\
        Computer-based Science and Technology Research Center, 20 Kavafi Str., 2121 Nicosia, Cyprus\\
        E-mail: \email{alexand@ucy.ac.cy}}

\abstract{We compute the mass spectrum for strange/charmed baryons in the 
partially quenched approach using $N_f=2$ twisted mass QCD 
configurations. We investigate two main issues: the size of 
lattice artefacts using three values of the lattice spacing (the smallest of
which is approximately 0.05 fm) and the dependence of baryon masses on meson
(or quark) masses. We thus perform a global fit in order to extrapolate 
simultaneously to the continuum limit and to the physical point. 
We estimate the masses of $\Omega_{sss}$, $\Xi_{dss}$, $\Lambda_{uds}$, 
$\Omega_{ccc}$, $\Xi_{dcc}$, $\Lambda_{udc}$.}

\FullConference{The XXVIII International Symposium on Lattice Field Theory, Lattice2010\\
		June 14-19, 2010\\
		Villasimius, Italy}

\begin{document}

\vspace*{-1.5cm}
\section{Introduction}

Simulations with two light degenerate sea quarks ($N_f=2$) and 
including also the strange sea quark ($N_f=2+1$) are nowadays 
standard. The ETM Collaboration has generated a substantial sample of 
$N_f=2$ ensembles 
at four values of the lattice spacing (ranging from 0.1 to 0.05 fm), several
values of the light sea quark mass and several physical volumes.
Using these ensembles one can study the cut-off effects on observables 
and the insight gained provides valuable input for the choice of parameters
for the $N_f=2+1+1$ (i.e. including both the strange and the charm sea quarks) 
simulations under production. Preliminary results using $N_f=2+1+1$
simulations have been presented at this conference~\cite{gregorio}.

In the present study we therefore use $N_f=2$ ensembles with a partially 
quenched setup in which the strange and charm quarks are added only as
valence quarks. For heavy quarks the compton wavelength of the 
associated heavy-light meson is small 
compared to present attainable lattice spacings which means that cut-off 
effects can be large. The charm quark mass is at the upper boundary 
of the range of masses that can be simulated at present for the 
coarsest lattice spacing used in the continuum limit extrapolation 
($a\sim 0.1fm$ for which $m_c a\lesssim 1$). In order to safely control 
this extrapolation it is thus important to asses the size of lattice 
artefacts affecting the observables of interest. 

Our goal is to extend the study of Ref.~\cite{Alexandrou:2009qu} by including 
a finer lattice spacing $a\simeq0.051$ fm. We would like, in addition, to
compute the low-lying spectrum of strange and charmed baryons. 
In this contribution we present 
preliminary results for the masses of the strange baryons 
$\Omega_{sss}$, $\Xi_{dss}$, $\Lambda_{uds}$ and the corresponding 
charmed baryon obtained by substituting the strange quark with 
the charm quark ($\Omega_{ccc}$, $\Xi_{dcc}$, $\Lambda_{udc}$).
Preliminary results for the low-lying strange baryon spectrum with 
$N_f=2+1+1$ gauge configurations~\cite{drach} and for the spectrum of 
static-light baryons with $N_f=2$ configurations~\cite{Wagner:2010hj} 
have also been presented at this conference.

\vspace*{-0.3cm}
\section{Setup}

The lattice discretization used for the doublet of degenerate light
quarks is Wilson twisted mass QCD at maximal twist~\cite{Frezzotti:2000nk} 
whose action reads (in the twisted basis)
\beq
S_{\rm{light}}^{\rm tmQCD} = a^4 \sum_{x}  \bar{\chi}_l(x) \,
\Big( \frac{\gamma_\mu}{2} (\nabla_\mu+\nabla_\mu^*) - \frac{a}{2} \nabla_\mu^{*}\nabla_\mu + M_{cr} +
i \gamma_5 ~\tau_3~\mu_l  \Big) \, \chi_l(x)
\eeq
where $\nabla_\mu$, $\nabla_\mu^{*}$ are forward and backward covariant
derivatives, $M_{cr}$ is the Wilson critical mass and $\mu_l$ is 
the light quark mass. 

The strange and charm (which in the following are referred to as ``heavy'') 
quarks are added 
here only as valence quarks {\it \`a la} 
Osterwalder-Seiler and their action reads
\beq
S_{\rm heavy}^{\rm{OS}} = a^4 \sum_{x} \sum_{h=s}^{c} \bar{\chi}_h(x) \,
\Big( \frac{\gamma_\mu}{2}(\nabla_\mu+\nabla_\mu^*) - \frac{a}{2}
\nabla_\mu^{*}\nabla_\mu + M_{cr} + i \gamma_5 ~\mu_h  \Big) \, \chi_h(x)
\eeq
where $\mu_s$ and $\mu_c$ are the strange and charm (valence) quark masses.
In order to remove the determinant of the strange and charm quarks, ghosts have
to be added correspondingly. Concerning the gauge actions, ETMC uses
the tree-level Symanzik improved gauge action.   

The main advantage of this regularization with respect to the
standard Wilson one is that the spectrum and the matrix 
elements extracted from correlation functions are automatically $O(a)$
improved~\cite{Frezzotti:2003ni}.
The drawback is that parity and isospin are explicitly broken by $O(a^2)$ 
lattice artefacts and are recovered only in the continuum limit. Here we 
use ETMC configurations generated at three values of the lattice spacing
$a\in\{0.051,0.064,0.080\}$ fm and physical volumes $L\sim2.0\div2.4$
fm (the scale has been set through $f_\pi$ in Ref.~\cite{Baron:2009wt}).
Light sea quark masses correspond to pion masses $M_{\pi}\in[290,520]$ MeV
while partially quenched valence strange/charm quarks correspond 
to heavy-light meson masses
$M_K\in[520,710]$ MeV and $M_D\in[1.80,2.40]$ GeV. In all we have 
40 different combinations ($M_\pi$,$M_{hl}$). In order to combine data at
different lattice spacings we express the value of the masses in units 
of $r_0$~\cite{Sommer:1993ce}. 
For the three lattice spacings considered here the values  
$r_0/a \in\{8.36, 6.73, 5.36\}$ are taken from Ref.~\cite{Baron:2009wt} 

\vspace*{-0.3cm}
\section{Numerical results}

\begin{figure}[hbt] 
\vspace*{-0.5cm}
\centerline{\includegraphics[width=10.9cm]{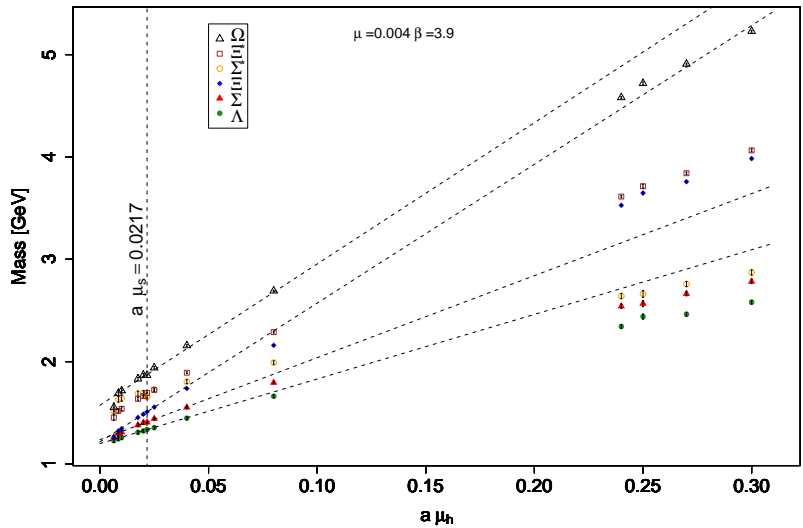}}
\vspace*{-0.3cm}
\caption{The dependence of the octet and decuplet baryon masses on 
$\mu_h$. Dashed lines corespond to linear fits performed in the strange region.}
\label{fig1} 
\end{figure}

An important issue in our study is the dependence of the baryon
masses upon the ``heavy'' quark mass $\mu_h$ in the strange and in the 
charm region. At $a=0.080$ fm and $M_\pi\simeq340$ MeV this 
dependence is shown in Fig.~\ref{fig1}.

From Fig.~\ref{fig1} we observe
that baryon masses depend linearly on $\mu_h$ 
both in the strange and in the charm region but with two different slopes.
This behavior will be further discussed in what follows. 
For what concern meson masses, in the case of the Kaon we observe a dependence 
$M_K^2 \propto \mu_h$, in agreement with the fact that the Kaon can still be 
considered a pseudo Goldstone boson. For the $D$ meson instead
we observe a dependence $M_D\propto \mu_h$ as predicted by heavy quark 
effective theory (HQET), with no evidence of $1/\mu_h$ term.   
In the following we will consider the functional dependence of baryon masses 
upon $M_\pi$ and $M_{hl}$ because this allows to extrapolate to the
physical point without knowing the values of the renormalized quark masses.
The observations above imply that baryon masses depend quadratically on 
$M_K$ in the strange region while depend linearly on $M_D$ in the charm 
region.
 
\begin{figure}[!t] 
\vspace*{-0.8cm}
\centerline{\includegraphics[width=10.9cm]{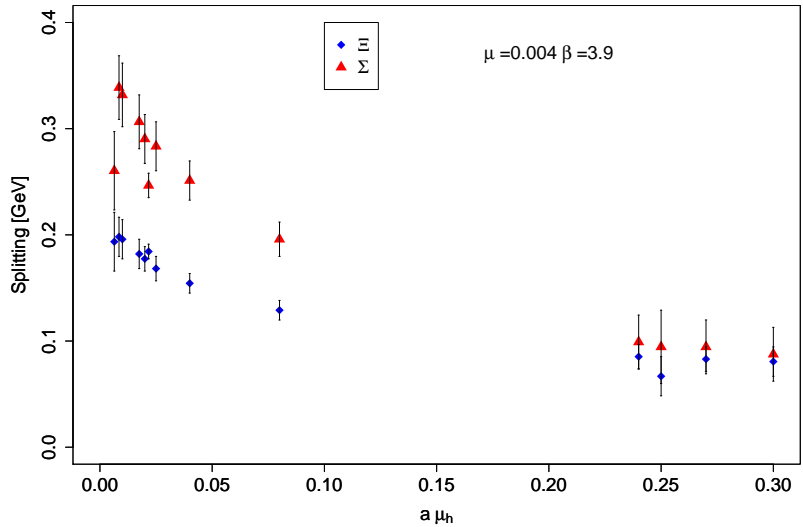}}
\vspace*{-0.3cm}
\caption{ $\Sigma/\Sigma^*$ and $\Xi/\Xi^*$ splittings as function
  of $\mu_h$.}
\vspace*{-0.1cm}
\label{fig2} 
\end{figure}

From Fig.~\ref{fig1} and Fig.~\ref{fig2} it is also evident that the 
splitting between 
$J=1/2$ and $J=3/2$ states ($\Sigma/\Sigma^*$ and $\Xi/\Xi^*$) 
clearly diminishes with the increase of $\mu_h$. In quark models, 
this observation is explained thanks to the fact that the spin-spin coupling 
part of the $q-q$ potential is inversely proportional to the masses of the two 
quarks $\frac{{\bf s}_i\cdot {\bf s}_j}{\mu_i \mu_j}$. In HQET,
the splitting of baryons containing one heavy quark (e.g. the $\Sigma_{udh}/\Sigma_{udh}^*$) is proportional to $1/\mu_h$.

Hadron masses $M_H$ are extracted from the two point correlators
$C_{\rm H}(t)=\sum_{\bf x}\langle H(t,{\bf x}) H^\dag(0,{\bf
  0})\rangle$ of the corresponding interpolating operators $H$ at large 
time distances. The interpolating operators $H$ are those
of Ref.~\cite{Alexandrou:2009qu} and to improve their overlap with the ground
state we apply Gaussian smearing and use APE smearing for the links 
that enter the hopping function. 
At large Euclidean time separation the value of
the hadron mass can be extracted by fitting the effective mass defined by 
$M_{\rm H}^{\rm eff}(t)=\frac{1}{a}\ln\frac{C_{\rm H}(t)}{C_{\rm H}(t+a)}$ 
to a constant. 

\begin{figure}[!b]
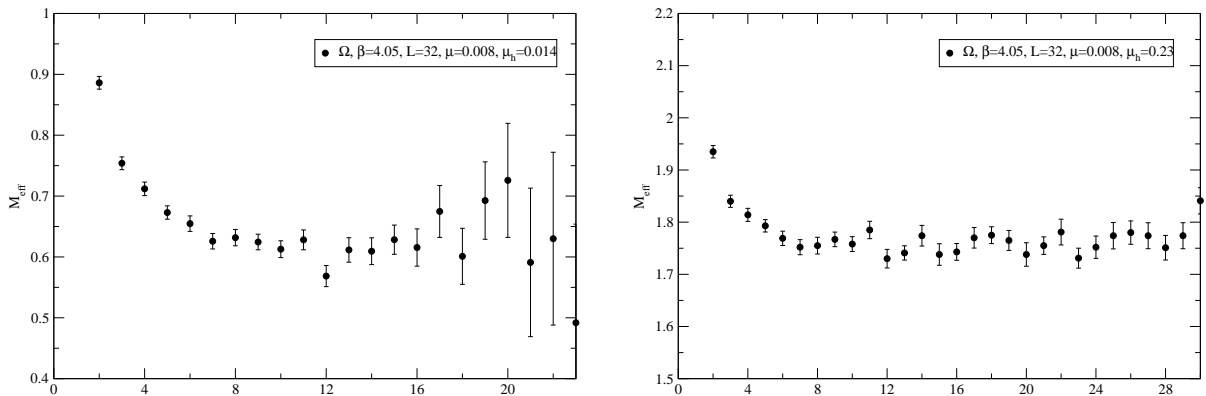
 
\vspace*{0.3cm}
\hspace*{-0.7cm}
\includegraphics[width=7.6cm]{./Figures/meff_Omega_mu0.008_muh0.014_L32_b4.05.eps}
\put(20,00){\includegraphics[width=7.6cm]{./Figures/meff_Omega_mu0.008_muh0.23_L32_b4.05.eps}}
\vspace*{-0.0cm}
\caption{Left: $M_{\Omega_{sss}}^{\rm eff}(t)$. Right:  $M_{\Omega_{ccc}}^{\rm eff}(t)$.}
\vspace*{-0.4cm}
\label{fig3} 
\end{figure}

It turns out that the statistical error on $M_{\rm H}^{\rm eff}(t)$ 
for the strange baryons grows faster in time than in the case of the 
charmed baryons. In the case of the $\Omega_{sss}$ and $\Omega_{ccc}$ 
this is illustrated in Fig.~\ref{fig3}. It is easy to show that 
the statistical error on
$M_{\Omega_{hhh}}^{\rm eff}(t)$ is
\beq
\Delta M_{\Omega_{hhh}}^{\rm eff}(t)\propto \exp(M_{\Omega_{hhh}}-\frac{3}{2}M_{\bar{h}h})t
\eeq
where $M_{\bar{h}h}$ is the mass of the $\bar{h}h$ meson made of an heavy
and an anti-heavy quark. This phenomenon is then probably explained by the
fact that the gap $\Delta_{\Omega_{ccc}}\equiv
M_{\Omega_{ccc}}-\frac{3}{2}M_{\bar{c}c}$ has a smaller value 
than $\Delta_{\Omega_{sss}} \equiv M_{\Omega_{sss}}-\frac{3}{2} M_{\bar{s}s}$.
At the physical point  
$M_{\Omega_{sss}}=1672$ MeV while the unphysical $\bar s s\equiv\eta_s$ meson
would have a mass $M_{\bar s s}\approx\sqrt{2M^2_K-M^2_\pi}\approx 690$ 
MeV~\cite{Davies:2009tsa} and the gap $\Delta_{\Omega_{sss}}\approx 640$ MeV. 
In the charm
case instead, the preliminary prediction from the present work gives
$M_{\Omega_{ccc}}\approx4730$ MeV while the $\bar c c$ meson can be identified 
with the $\eta_c$ meson which has a mass $M_{\eta_c}=2980$
MeV. The gap $\Delta_{\Omega_{ccc}}\approx 260$ MeV is
therefore sensibly smaller than in the strange case. Presumably, this fact 
remains true for the values of the meson masses we have in our simulations 
but we still need to check numerically this conjecture.

\vspace*{-0.3cm}
\subsection{$\Omega_{sss}$ and $\Omega_{ccc}$}

\begin{figure}[!t] 
\vspace*{-1.0cm}
\centerline{\includegraphics[width=11.0cm]{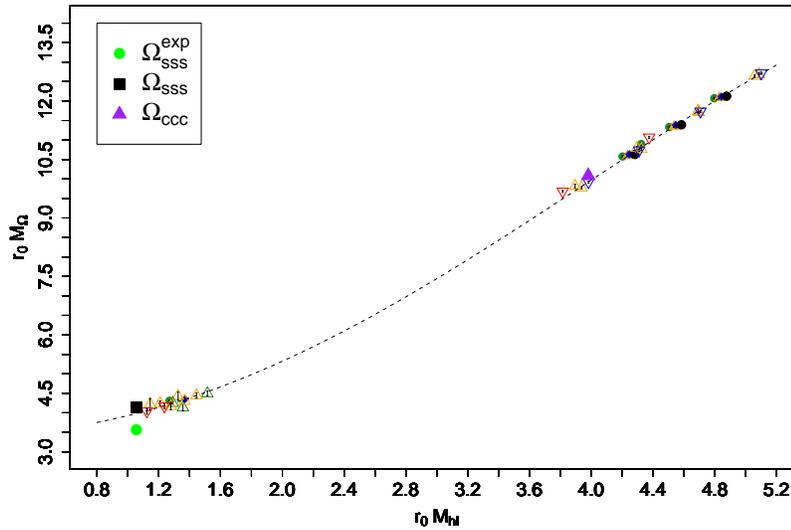}}
\vspace*{-0.7cm}
\caption{ $M_{\Omega}$ as function of $M_{hl}$. The dashed line is
an interpolating form between the $s$ and $c$ region.}
\vspace*{-0.3cm}
\label{fig4} 
\end{figure} 

In Fig.~\ref{fig4} we present all the 40 data points for the $\Omega$ mass
fitted to a functional form which interpolates between the strange and 
the charm region. This plot already shows the smallness of lattice artefacts. 
The functional form reduces, in the strange region, to the form
$M_\Omega=M_0+A M_\pi^2 +B M_{hl}^2$. In the charm region it reduces instead to
$M_\Omega= D+ E M_\pi^2 + F M_{hl}$. 
The two forms fit well the data: using 13 data points in the strange region we
obtain $\chi^2_{\rm d.o.f.}=1.56$; using 27 data points in the charm region we
find $\chi^2_{\rm d.o.f.}=1.15$. Lattice artefacts are visible in the strange
region and the inclusion of a term $A_0 a^2$ to the functional form 
above lower the $\chi^2_{\rm d.o.f.}$ from 1.56 to 0.92. 
For charmed baryons we do not see any cut-off effect.  

\begin{figure}[!t] 
\vspace*{-0.8cm}
\centerline{\includegraphics[width=11.0cm]{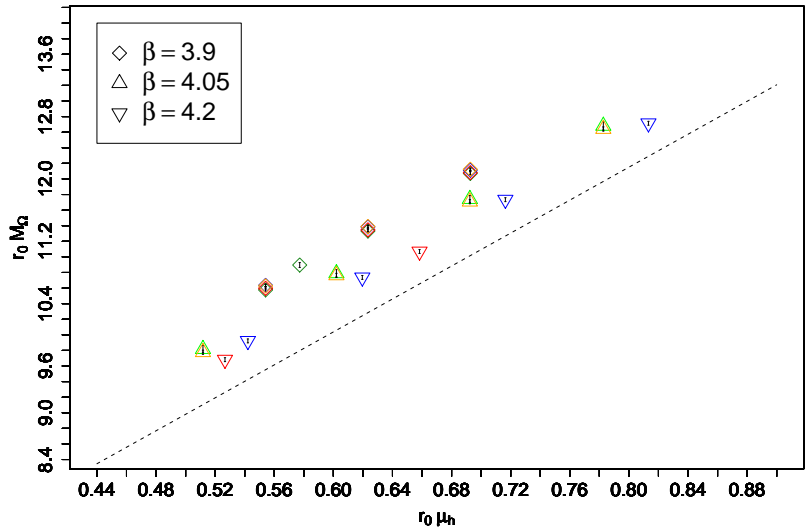}}
\vspace*{-0.3cm}
\vspace*{-0.3cm}
\caption{ $M_{\Omega}$ as function of $\mu_h$ in the charm region. The
  dashed line is the continuum limit obtained by plotting the fitting function
 described in the text after extrapolating to $a=0$.}
\vspace*{-0.3cm}
\label{fig5} 
\end{figure} 

This is due however to the choice of studying the behaviour of baryon masses
as function of meson masses. Had we chosen to study their dependence 
upon the renormalized quark masses $\mu_l$ and $\mu_h$ (obtained from the bare
masses by multiplying them by $Z_\mu$ taken from Ref.~\cite{Blossier:2010cr}) 
we would have immediately remarked the presence of lattice artefacts, at least
in the charm region. In this region, a fit to the form 
$M_\Omega= D+ E \mu_l + F \mu_h$ (i.e. not including lattice artefacts) is not
sufficient and gives a huge $\chi^2_{\rm d.o.f.}$. In order to obtain a 
reasonable $\chi^2_{\rm d.o.f.}=1.18$ one needs to add lattice artefacts 
(both $\mu_h$-independent and $\mu_h$-dependent). 
Fig.~\ref{fig5} shows the data points together with the curve obtained by 
plotting the fitting function after setting $a=0$. 

Lattice artefacts are instead hardly visible in the strange region. 
Here, a fit to the form $M_\Omega= A+ B \mu_l + B \mu_h$ gives a
reasonable $\chi^2_{\rm d.o.f.}=1.39$ and adding lattice artefacts (dependent 
or independent on the quark masses) does not improve the fit.

We remark that $M_\Omega$ depends very mildly on $M_\pi$ and therefore the
extrapolation to the physical $M_\pi$ seems not to pose any problem.
By interpolating also to the physical value of $M_K$ we get the result
$M_{\Omega_{sss}}=1.86(20)$ GeV which is consistent with the analysis 
in Ref.~\cite{Alexandrou:2009qu} but still $10\%$ larger than the experimental value.
Due to the previous considerations and the analysis performed, 
this discrepancy seems not to be related 
to the continuum limit extrapolation or to the 
extrapolation in the light quark mass. Extrapolation to the 
physical $(M_\pi,M_D)$ point gives the prediction 
$M_{\Omega_{ccc}}=4.73(40)$ GeV (the experimental value is not known) and the
ratio $M_{\Omega_{sss}}/M_{\Omega_{ccc}}=0.393(54)$

\vspace*{-0.0cm}
\subsection{$\Lambda_{uds}$ and $\Lambda_{udc}$}

In the case of the $\Lambda$ baryon, the dependence on $M_\pi$ is much
stronger than in the previous case and the inclusion of the term 
proportional to $M_\pi^3$ is crucial
and reduces the $\chi^2_{\rm d.o.f.}$ of a factor $\sim 0.5$ in both
the strange and the charm region. Lattice artefacts are hardly visible and 
the functional forms we have used to fit are 
$M_\Lambda=M_0+ A M_\pi^2 + B M_{hl}^2 + C M_\pi^3$ (in the
strange region) and $M_\Lambda=D + E M_\pi^2 + F M_{hl} + G M_\pi^3$ (in the
charm region). Of course the inclusion of chiral logarithms would affect
the extrapolation to the physical point. For these preliminary results we 
have however neglected them and performed only a rough fit using the 
forms written above.


By extrapolating to the physical $(M_\pi,M_K)$ point we obtain
$M_{\Lambda_{uds}}=1.20(10)$ GeV which has to be compared with the 
experimental value $M^{\rm exp}_{\Lambda_{uds}}=1.116$ GeV. 
By extrapolating to the physical 
$(M_\pi,M_D)$ point we have $M_{\Lambda_{udc}}=2.24(18)$ GeV which is
in good agreement with the experimental value $M^{\rm exp}_{\Lambda_{udc}}=2.286$ 
GeV. 

\vspace*{-0.0cm}
\subsection{$\Xi_{dss}$ and $\Xi_{dcc}$}

Twisted mass QCD breaks explicitly isospin symmetry and thus $\Xi_{uss}^0$ and 
$\Xi_{dss}^-$ (or equivalently $\Xi_{ucc}^{++}$ and $\Xi_{dcc}^+$) are not 
degenerate. We thus preform a combined fit of both $\Xi_{uss}^0$ and 
$\Xi_{dss}^-$ data with the form $M_{\Xi^{\{0,-\}}}=M_0+A M_\pi^2 
+B M_{hl}^2 +C M_\pi^3 + A_{\{0,-\}}a^2$ where the coefficients  $A_{\{0,-\}}$ are
different for the two sets of data. Analogously we perform a combined fit 
of both $\Xi_{ucc}^{++}$ and $\Xi_{dcc}^+$ data with the form 
$M_{\Xi^{\{++,+\}}}=D+E M_\pi^2 +F M_{hl}+G M_\pi^3
 + D_{\{++,+\}}a^2$. The dependence on $M_\pi$ of the mass of the doubly
 charmed $\Xi$ turns out to be considerably less pronounced
 than for the standard (strange) $\Xi$. The coefficient $A_-$ is substantially
 smaller than $A_0$ and in the charm case $D_+$ is compatible with zero and
 can be removed from the fit function.  
Fits work well and in the continuum limit, at the physical point,
we get $M_{\Xi_{dss}}=1.37(12)$ GeV (to be compared with 
$M^{\rm exp}_{\Xi_{dss}}=1.32$ GeV) and $M_{\Xi_{dcc}}=3.52(25)$ GeV 
(in perfect agreement with $M^{\rm exp}_{\Xi_{dcc}}=3.52$ GeV).

\vspace*{-0.3cm}
\section{Conclusions}

In this preliminary study we have shown that, when baryon masses are 
analyzed as function of meson masses, lattice artefacts are always small 
and in some cases (notably in the charm region) hardly visible. 
They are instead clearly visible when baryon masses are analyzed as function 
of quark masses. As expected lattice artefacts are larger in the charm
region, where they increase proportionally to $\mu_h$. 
The chiral extrapolation in the light quarks confirms 
to be critical and a term of order $M_\pi^3$ is needed for both $\Xi$ and 
$\Lambda$ (it is particularly evident in this last case). 
$M_{\Omega_{sss}}$ is still $10\%$ larger than the experimental value 
and the source
of this discrepancy seems not to be related to the continuum limit 
extrapolation or to the extrapolation in the light quark mass. Further
investigations are needed to clarify this issue. Results for $M_{\Xi_{dss}}$,
$M_{\Lambda_{uds}}$, $M_{\Xi_{dcc}}$ and $M_{\Lambda_{udc}}$ nicely agree 
with the experimental values. 
We have moreover obtained a prediction for $M_{\Omega_{ccc}}=4.73(40)$ GeV. 
We are computing all the correlation functions needed to extract the whole 
low-lying spectrum of strange/charmed baryons. 
A complete analysis, including a more careful assessment of both statistical
and systematic errors, will be performed in the near future. 

\vspace*{-0.3cm}
\section*{Acknowledgements}

The computer time for this project was made available to us by CNRS 
on the BlueGene system at GENCI-IDRIS (Grant 2010-052271) and CCIN2P3 
in Lyon. We thank these computer centers and their staff for all 
technical advice and help. 
M.~Papinutto acknowledges financial support by a Marie Curie European
Reintegration Grant of the 7th European Community Framework Programme 
under contract number PERG05-GA-2009-249309.


\end{document}